\documentclass[prl,amsmath,amssymb,superscriptaddress,twocolumn]{revtex4-2}

\usepackage{graphicx}
\usepackage{dcolumn}
\usepackage{epsfig}
\usepackage{float}

\usepackage[bbgreekl]{mathbbol}

\usepackage{mathrsfs}

\usepackage[cal=boondox,scr=boondoxo]{mathalfa}
\def\mycal#1{\mathscr{#1}}

\usepackage{trsym} 
\def\psym{\TransformVert}
\def\hsym{\InversTransformVert}

\def\bra#1{\mathinner{\langle{#1}|}}
\def\ket#1{\mathinner{|{#1}\rangle}}
\def\inner#1#2{\mathinner{\langle{#1}|{#2}\rangle}}
\def\braket#1{\mathinner{\langle{#1}\rangle}}
\def\sandwich#1#2#3{\mathinner{\langle{#1}|{#2}|{#3}\rangle}}

\def\id{\mathbb{1}} 

\def\Texp{\mathrm{T}\hspace{-1mm}\exp}

\def\nodag{^{\vphantom{\dag}}}

\usepackage[T3,T1]{fontenc}
\DeclareSymbolFont{tipa}{T3}{cmr}{m}{n}
\DeclareMathAccent{\invbreve}{\mathalpha}{tipa}{16}

\usepackage{accents}
\newlength{\hhatheight}

\usepackage{xcolor}

\def\highlight#1{{#1}}

\makeatother

\begin{document}

\title{
Floquet Gauge Pumps as Sensors for Spectral Degeneracies Protected by \\ Symmetry or Topology
}

\author{Abhishek Kumar}
\affiliation{Department of Physics, Indiana University, Bloomington, Indiana 47405, USA}

\author{Gerardo Ortiz}
\affiliation{Department of Physics, Indiana University, Bloomington, Indiana 47405, USA}
\affiliation{Quantum Science and Engineering Center, Indiana University, Bloomington, Indiana 47405, USA}

\author{Philip Richerme}
\affiliation{Department of Physics, Indiana University, Bloomington, Indiana 47405, USA}
\affiliation{Quantum Science and Engineering Center, Indiana University, Bloomington, Indiana 47405, USA}

\author{Babak Seradjeh}
\affiliation{Department of Physics, Indiana University, Bloomington, Indiana 47405, USA}
\affiliation{Quantum Science and Engineering Center, Indiana University, Bloomington, Indiana 47405, USA}
\affiliation{IU Center for Spacetime Symmetries, Indiana University, Bloomington, Indiana 47405, USA}

\begin{abstract}
    We introduce the concept of a Floquet gauge pump whereby a dynamically engineered Floquet Hamiltonian is employed to reveal the inherent degeneracy of the ground state in interacting systems. We demonstrate this concept in a one-dimensional XY model with periodically driven couplings and transverse field. In the high-frequency limit, we obtain the Floquet Hamiltonian consisting of the static XY and dynamically generated Dzyaloshinsky-Moriya interaction (DMI) terms.  The dynamically generated magnetization current depends on the phases of complex coupling terms, with the XY interaction as the real and DMI as the imaginary part. As these phases are cycled, the current reveals the ground-state degeneracies that distinguish the ordered and disordered phases. We discuss experimental requirements needed to realize the Floquet gauge pump in a synthetic quantum spin system of interacting trapped ions.
\end{abstract}

\date{\today}

{
\let\clearpage\relax
\maketitle
}





\emph{Introduction.}---%
The nontrivial topology of gapped phases of matter is often manifested in states or modes localized at the boundary or defects of the system protected by symmetries and the bulk topological gap. This bulk-boundary correspondence has been rigorously proven in certain cases, especially in non-interacting systems~\cite{Schnyder_2008,Kitaev_2009,Teo_2010,Ryu_2010,Fu_2011,Isaev_2011,Prodan_2016}. Thus, the presence and properties of boundary modes is used as a telltale experimental signature of bulk topology~\cite{Law_2009}. However, boundary modes can arise in many other, topologically trivial cases as well~\cite{Liu_2012}. Moreover, the structure of boundary modes is far from clear in generic interacting many-body systems~\cite{Goldstein_2012,Ortiz_2014,Kells_2015,Ortiz_2016}. Therefore, more robust probes of bulk topology are highly sought after~\cite{Grosfeld_2011,Hassler_2010,Grosfeld_2011a,Bose_2011,Alicea_2012,Dahan_2017}.

An example is provided by the fractional Josephson current, $J_s$, between topological superconductors supporting Majorana bound states as the phase difference $\Delta\phi$ between the superconductors is cycled by changing the magnetic flux enclosed by the system~\cite{Kitaev_2001,Kwon_2003,Fu_2009,Rokhinson_2012}. Unlike the Cooper-pair-mediated conventional Josephson current between trivial superconductors $J_s \propto \sin \Delta\phi$, the fractional Josephson current is dominated by quasiparticle tunneling through the Majorana bound states, $J_s\propto \sin(\Delta\phi/2)$. In the presence of interactions, the fractional Josephson current probes the topological degeneracy of the interacting ground state in a given fermion parity sector~\cite{Ortiz_2014}.

In this Letter, we consider a general spatially resolved probe provided by the variations in the current flowing through the bridge between two gapped phases as a relevant gauge field is varied in a cycle. Ground-state degeneracies through this cycle produce an anomalous periodicity of the corresponding current on the cycle parameters\highlight{~\cite{Cobanera_2014,Cobanera_2016}}. As energy is pumped through the junction, we call such probes ``gauge pumps.'' In a topological phase, degeneracies are  produced by topological boundary modes localized at the bridge. However, it is important to note that the role of bulk topology here is to guarantee the existence of degeneracies in the many-body spectrum. While the notion of topology depends on the choice of the local basis, or language, the existence of a many-body degeneracy is independent of this representation. Thus, depending on the local basis, gauge pumps can detect ground-state degeneracies of bulk topological or spontaneous symmetry broken phases.

 We extend the notion of gauge pumps to periodically driven systems. We introduce the ``{Floquet gauge pump}'' realized by a periodic drive protocol that both imprints and controls the bridge geometry. As a concrete demonstration, we study the driven XY model in transverse field and show how Dzyaloshinsky-Moriya interaction (DMI) terms can be dynamically generated and tuned by the periodic drive. Upon Jordan-Wigner transformation \cite{Jordan_1928}, the DMI and exchange couplings map to complex fermion hopping and pairing amplitudes that realize trivial and topological superconducting phases of fermions corresponding, respectively, to disordered and ordered phases of the original spins. The spatial profile of the drive can be used both to create the bridge geometry of a gauge pump and to cycle its gauge parameters~\footnote{
 For other examples of spatial variation in a Floquet drive, see Refs.~\cite{Katan_2013,Kundu_2016,Rodriguez-Vega_2019}.
 }. In the original spin model, the gauge current corresponds to the rate of change of magnetization. Thus, many-body degeneracies are revealed by an anomalous dependence of magnetization current on drive parameters.
The gauge pump and its Floquet realization proposed in this work offer a powerful and widely useful probe of symmetry-protected degeneracies of topological and ordered phases of quantum matter.

\emph{Floquet gauge pump.}---%
A gauge pump is realized by a cyclic Hamiltonian $H(\phi)$ in the gauge parameters $\phi$~\footnote{
By contrast, a topological quantum pump such as the Thouless pump~\cite{Thouless_1983} is realized by an adiabatic cycle of the Hamiltonian that breaks the symmetry underlying the topological phase and pumps a quantized number of topological boundary modes through the system, resulting in a current flow.
}. A constant $\phi$ can be gauged away as $U_g^\dagger(\phi) H(\phi) U\nodag_g(\phi) = H(0)$ with a unitary gauge transformation $U_g(\phi)$. The gauge pump is constructed by ``bridging'' two such Hamiltonians, $H_L$ and $H_R$, to form $H_\text{gp} = H_L(\phi_L)\otimes\id_R + H_{LR} + \id_L\otimes H_R(\phi_R)$, where $\id_{L(R)}$ is the identity operator on the left (right) side of the bridge given by $H_{LR}$. After a gauge transformation $U_{g} = U_{gL}(\phi_L)\otimes U_{gR}(\phi_R)$, we have the gauge-equivalent Hamiltonian $U_{g}^\dagger H_\text{gp} U\nodag_{g} = H_L(0)\otimes \id_R + H_b(\phi_h,\phi_p) + \id_L \otimes H_R(0)$, where $\phi_h \equiv \phi_L - \phi_R$, $\phi_p \equiv \phi_L+\phi_R$ , and $H_b = U_{g}^\dagger H\nodag_{LR} U\nodag_{g}$. Then, the gauge currents on each side are~\cite{SM}
\begin{align}
j_L &:= \left< \frac{\partial H_\text{gp} } {\partial\phi_L } \right> = \left< \frac{\partial H_b }{\partial\phi_p} \right>_g + \left< \frac{\partial H_b }{\partial\phi_h} \right>_g , \\ 
j_R &:= \left< \frac{\partial H_\text{gp} } {\partial\phi_R } \right> = \left< \frac{\partial H_b }{\partial\phi_p} \right>_g - \left< \frac{\partial H_b }{\partial\phi_h} \right>_g,
\end{align}
where the expectation values $\braket{\cdots}_g = \braket{U_g\nodag \cdots U_g^\dagger}$.
Note that the current flow is set with respect to the bridge, so the positive values have opposite directions on each side.
If the bridge Hamiltonian $H_b$ is a function of $\phi_h$ only, the two currents $j_L = - j_R$ and no gauge charge is accumulated in the bridge itself. However, if the bridge Hamiltonian also depends on $\phi_p$, then $j_b := j_L + j_R = 2\braket{\partial H_b/\partial\phi_p}_g$ must be carried by the bridge itself. If the bridge is ``grounded,'' e.g. in a transport geometry of a mesoscopic device, this can flow through the bridge. Otherwise, the gauge charge will accumulate in the bridge.

The Floquet gauge pump provides two complementary functions. First, as we show, periodic drive protocols with spatial variations may be used both to engineer the gauge parameters and to imprint the gauge pump geometry~\cite{Katan_2013,Kundu_2016,Rodriguez-Vega_2019}. Second, drive parameters can be tuned to engineer Floquet topological phases of the system\highlight{~\cite{Oka_2009,Wang_2013,Gomez-Leon_2014,Poudel_2015,McIver_2019,Topp_2019,Li_2020,Katz_2020,Rodriguez-Vega_2020,Ozawa_2019,Oka_2019,Rudner_2020a,Rudner_2020}}. The signatures of both equilibrium and Floquet topological phases can then be probed by the dependence of the gauge current pumped through the system on tunable gauge parameters. To illustrate this, note that the stroboscopic dynamics of the driven Hamiltonian $H_\text{gp}(t) = H_L(t)\otimes\id_R + H_{LR}(t) + \id_L\otimes H_R(t) = H_\text{gp}(t+2\pi/\Omega)$ with drive frequency $\Omega$ is governed by the Floquet Hamiltonian~\cite{Floquet_1883,Shirley_1965,Sambe_1973} $H_\text{gp}^\text{F} = i(\Omega/2\pi) \ln \Texp[-i\oint H_\text{gp}(t)dt] \equiv H_{L}^\text{F} \otimes \id_R + H_{LR}^\text{F} + \id_L\otimes H_{R}^\text{F}$, where the time-ordered exponential is over a full cycle of the drive and we have set Planck's constant $\hbar=1$. At high frequency up to $\mathcal{O}(\Omega^{-2})$, we may expand~\cite{Eckardt_2015,Rodriguez-Vega_2018} $H_{a}^\text{F} = H_{a}^{(0)} + \sum_{n\in\mathbb{N}} [H_{a}^{(-n)}, H_{a}^{(n)}]/(n\Omega)$ for $a=L,R$ and
\begin{align}
H_{LR}^\text{F} 
	= H_{LR}^{(0)} 	&+ \sum_{n\in\mathbb{N}} \left\{ \frac{[H_{LR}^{(-n)}, H_{LR}^{(n)}]}{n\Omega} \right. \nonumber\\
						&+ \left. \sum_a\left( \frac{[H_{LR}^{(-n)}, H_{a}^{(n)}]}{n\Omega} + \text{H.c.} \right)\right\},
\end{align}
where $O^{(n)} = (\Omega/2\pi) \oint O(t) e^{in\Omega t} dt$ are the Fourier components  of operator $O$ and $\text{H.c.}$ is the Hermitian conjugate.
We denote the gauge parameters in this Floquet Hamiltonian as $\phi_L(\lambda)$ and $\phi_R(\lambda)$ with $\lambda$ denoting the drive parameters, such as frequency, harmonic amplitudes, and phases. The two-fold function of the Floquet gauge pump is then to provide independent realizations and tuning of $\phi_{L}$ and $\phi_R$. Figure~\ref{fig:sketch} sketches our setup.

\begin{figure}[t]
    \begin{center}
        \includegraphics[width=3.45in]{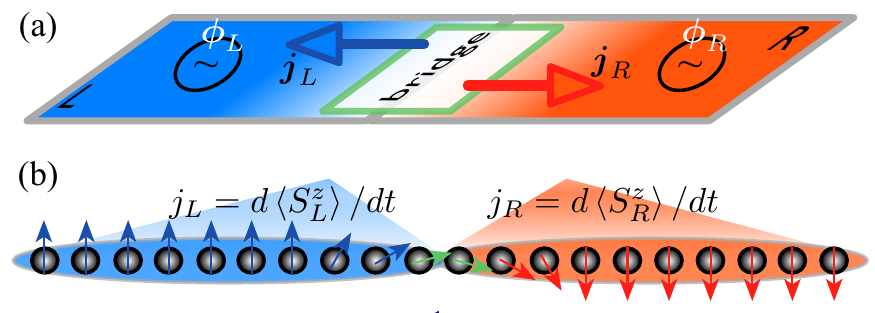}
    \end{center}
    \vspace{-0.7cm}
    \caption{Sketch of the Floquet gauge pump and its ion-trap realization. (a) By periodically driving the left ($L$) and right ($R$) sides of the system with different gauge parameters $\phi_L$ and $\phi_R$, the pump geometry and currents $j_L$ and $j_R$ can be engineered and controlled. (b) Trapped ions can realize a Floquet gauge pump with effective spin degrees of freedom, in which magnetization currents $d\langle S^z_{L,R}\rangle/dt$ can be pumped by controlling the drive protocols on each side.}
    \vspace{-0.5cm}
    \label{fig:sketch}
\end{figure}

The gauge currents $j_a(t) = \braket{\partial H_\text{gp}(t) / \partial \phi_a}$ are now time dependent. We show in the Supplemental Material~\cite{SM} that if the expectation value is calculated in Floquet modes $\ket{\Psi(t)} = e^{-i \epsilon t}\ket{\Phi(t)}$, where $\epsilon$ is the quasienergy and $\ket{\Phi(t+2\pi/\Omega)}=\ket{\Phi(t)}$ is the periodic eigenstate satisfying $[H_\text{gp}(t)-i\partial/\partial t]\ket{\Phi(t)} = \epsilon \ket{\Phi(t)}$, then 
the average gauge current is
\begin{align}
j_a^{(0)} \label{eq:ja0}
	= \frac{\partial \epsilon}{\partial \phi_a}.
\end{align}

The choice of physical state of the driven system requires care. Generic driven systems would, at infinitely long times, settle into a uniform mixed state by absorbing energy from the drive without bound~\cite{DAlessio_2014,Lazarides_2014},with the exception of integrable or many-body localized systems~\cite{Lazarides_2014a,Lazarides_2015,Ponte_2015a,Ponte_2015,Khemani_2016}. However, at intermediate times that can be extremely long for sufficiently large systems, the Floquet state describes the dynamics of the system rather well~\cite{Mori_2016,Kuwahara_2016,Abanin_2017,Weidinger_2017,Lindner_2017}. At high enough frequency, in particular, an initial equilibrium state of the average Hamiltonian $H_\text{gp}^{(0)}$ is nearly the same as the Floquet state. In the following, we will assume that this is indeed the case and calculate gauge currents from the Floquet spectrum.

\emph{Spin model}---%
To illustrate the concepts, here we consider a driven XY model in a transverse field,
$
H_\text{XY}(t)= \sum_{j} \left[ J_j^{x}(t)S^{x}_j S^{x}_{j+1} +J_j^{y}(t) S^{y}_j S^{y}_{j+1} + h_j^{z}(t) S^{z}_j \right], 
$
where $\{S^x_{j}, S^y_{j}, S^z_{j}\}$ are spin-$\frac{1}{2}$ operators, $\{J_j^x, J_j^y\}$ are nearest-neighbor couplings, and $h_j^z$ is the transverse field at lattice site $j$. We take the periodic drive to be independent and uniform on each side with $J_{j\in a}^{x,y}(t) = \overline J_a^{x,y} + \delta J^{x,y}_{a} \cos(\Omega t +\theta^J) $, and $h^{z}_{a}(t) = \overline h_a^z +  \delta h^{z}_{a} \cos(\Omega t +\theta^h_a)$.
With this choice, the high-frequency Floquet Hamiltonian takes the form~\cite{SM} $H_\text{XY}^\text{F} = H_{L}^\text{F} + H_{LR}^\text{F} + H_{R}^\text{F} +\mathcal{O}(\Omega^{-2})$,
\begin{align}
H_{a}^\text{F} 
	&= \overline H_{a} + \sum_{j\in a} \zeta_a \left(S^{x}_j S^{y}_{j+1} + S^{y}_j S^{x}_{j+1}\right) \\
H_{LR}^\text{F}
	&= \overline H_{LR} + \zeta_{LR} \left(S^{x}_{\mycal{l}} S^{y}_{\mycal{r}} + S^{y}_{\mycal{l}} S^{x}_{\mycal{r}}\right) \nonumber \\
    & \hspace{0.475in} + \xi_{LR} \left(S^{x}_{\mycal{l}} S^{y}_{\mycal{r}} - S^{y}_{\mycal{l}} S^{x}_{\mycal{r}}\right) ,
 \end{align}
Here $\overline H_a$ is the average Hamiltonian on side $a$ and $\zeta_a = - \delta J_a^{-}\delta h_{a}^z\sin(\theta_{a}^h-\theta^J) / \Omega$ with $\delta J_a^{\pm} = (\delta J_a^x \pm \delta J_a^y)/2$. At the junction connecting the sites $\mycal{l}\in L$ and $\mycal{r}\in R$, $\overline H_{LR}$ is the contribution from the averaged Hamiltonian, $\zeta_{LR} = - \delta J_L^{-} [\sin(\theta_{R}^h-\theta^J)\delta h_{R}^z + \sin(\theta_{L}^h-\theta^J)\delta h_{L}^z]/(2\Omega)$, and $\xi_{LR} = \delta J_L^{+} [\sin(\theta_{R}^h-\theta^J)\delta h_{R}^z - \sin(\theta_{L}^h-\theta^J)\delta h_{L}^z ] /(2\Omega)$ is the DMI term dynamically generated by the drive. The $z$ component of the magnetization current on side $a$ is defined as $j_a = d\braket{S^z_a}/dt$, where $S^z_a = \sum_{j\in a} S^z_j$.

It is mathematically convenient to analyze the spin gauge pump in the dual fermionic language. To this end, we  employ the Jordan-Wigner transformation~\cite{Jordan_1928} $S_j^x + i S_j^y = P\nodag_j \, c_j^{\dagger}$, $S_j^{z}= n_j -\frac{1}{2}$, with number operator $n_j=c_j^{\dagger}c\nodag_j$ ($c_j, c_j^\dagger$ are fermionic operators) at site $j$, and $P_j = \prod_{l<j} e^{i \pi n_l}$ as the fermion parity to the left of site $j$. This is followed by the gauge transformation $e^{i\phi_a} c^\dagger_j \to c^\dagger_j$ for $j\in a$, to find, up to a constant, the equivalent fermion Hamiltonian $\widetilde H_{L}^\text{F} + H_b + \widetilde H_{R}^\text{F}$,
\vspace{-0.5mm}
\begin{align}
\widetilde H_{a}^\text{F} 
	&= \sum_{j\in a} \left[ w_{a} c_j^{\dagger}c\nodag_{j+1} + \Delta_{a} c_j^{\dagger}c_{j+1}^{\dagger} + \mu_a n_j \right] + \text{H.c.}
	\label{eq:HFa} \\
H_b 
	&= w_b e^{i \phi_{h}} c_{\mycal{l}}^{\dagger}c\nodag_{\mycal{r}} + \Delta_b e^{i \phi_{p}}  c_{\mycal{l}}^{\dagger}c_{\mycal{r}}^{\dagger} +  \text{H.c.}, 
	\label{eq:Hb}
\end{align}
where chemical potential $\mu_a = \frac12 h_a^z$,  hopping amplitudes $w_a = \frac12 J_a^+$ and $w_b = \frac12 |(J_L^+ + i \xi_{LR})|$, and pairing amplitudes $\Delta_a = \frac12 |(J_a^- + i\zeta_a)|$ and $\Delta_b = \frac12 |(J_L^- + i \zeta_{LR})|$, $J_a^\pm = (\overline J_a^x \pm \overline J_a^y)/2$ are all real, and
\begin{align}
\phi_h &= \phi_{hb} + (\phi_{L}-\phi_{R}), 
\\
\phi_p &= \phi_{pb} + (\phi_{L}+\phi_{R}), 
\end{align}
with $\phi_a = \frac12\arg(J_a^-+i\zeta_a)$ half of the pairing phase on each side, $\phi_{hb} = \arg(J_L^+ + i \xi_{LR})$ and $\phi_{pb} = \arg(J_L^- + i \zeta_{LR})$. This fermionic Hamiltonian is composed of a Kitaev chain~\cite{Kitaev_2001} on each side and a bridge Hamiltonian with both hopping and pairing terms at the junction, thus realizing an unconventional Josephson junction that can be controlled by the original drive parameters. In terms of fermions, the currents $j_a=d\braket{N_a}/d t$, where $N_a = \sum_{j\in a} n_j$ is the fermion number operator on side $a$. 

The Kitaev chain has two phases. For $|\mu_a| < 2 w_a$, there are unpaired Majorana bound states at the junction and the spectrum is doubly degenerate. In terms of spins, this corresponds to the ordered phase of the XY model. By contrast, when $|\mu_a| > 2w_a $, the spectrum is nondegenerate and corresponds to a disordered spin chain. The presence or absence of Majorana fermions at the junction affects the current across the junction.

\emph{Floquet gauge current}---%
The average Floquet gauge current is given by Eq.~(\ref{eq:ja0}), which requires the calculation of the quasienergies of $H_\text{XY}^\text{F}$. We will do so perturbatively in the bridge Hamiltonian, where the unperturbed system (without the bridge) has a Floquet spectrum $\ket{\Phi_\alpha}$ with quasienergies $\epsilon_\alpha$. The quasienergy of a state $\ket{\Phi_{\alpha=0}}$ in second-order perturbation theory is $\epsilon = \epsilon_0 + \epsilon_b$, where $\epsilon_b(\phi_h,\phi_p) = \sandwich{\Phi_0}{H_b}{\Phi_0} +\sum_{\alpha\neq0}|\sandwich{\Phi_0}{H_b}{\Phi_\alpha}|^2/(\epsilon_0-\epsilon_\alpha)$. We present the details of this calculation in Supplemental Material~\cite{SM} and quote the main results here.

In the trivial (disordered) phase, the first-order contribution vanishes since $H_b$ projects out of the unperturbed ground state. Then the current takes the form
\begin{align}
j_L^{(0)} &= F_\text{c}\sin(2\phi_p) + F_{L}\sin(\phi_p+\phi_h) + F_\text{t}\sin(2\phi_h), \hspace{-1mm} \\
j_R^{(0)} &= F_\text{c}\sin(2\phi_p) + F_{R}\sin(\phi_p-\phi_h) - F_\text{t}\sin(2\phi_h), \hspace{-1mm}
\end{align}
where $F_\text{c} \propto \Delta^2_b$, $F_{L,R}\propto w_b \Delta_b$, and $F_\text{t} \propto w^2_b$ are second order in bridge tunneling amplitudes. The parity of the nondegenerate state is fixed over the entire range of gauge parameters. 

\begin{figure}[t]
    \begin{center}
        \includegraphics[width=3.45in]{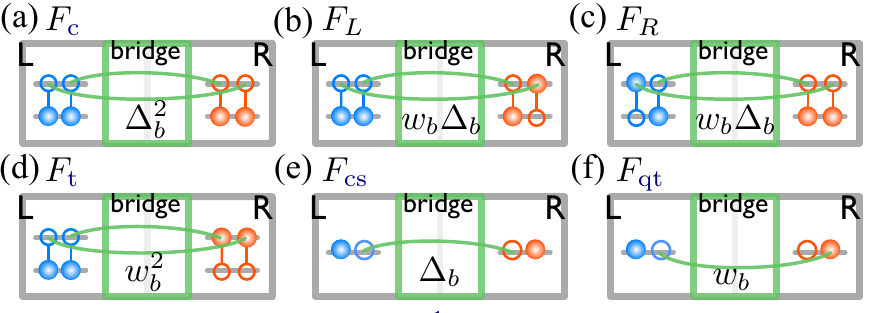}
    \end{center}
    \vspace{-0.7cm}
    \caption{Gauge current processes. In the trivial phase (a)-(d), second-order virtual processes due to pairing $\Delta_b c^\dagger_\mycal{l}c^\dagger_\mycal{r}$ (hopping $w_b c^\dagger_\mycal{l}c_\mycal{r}$ or $w_b c_\mycal{l}c^\dagger_\mycal{r}$) tunneling are shown by $\psym$ ($\psym$ or $\hsym$) in $L$ and $\psym$ ($\hsym$ or $\psym$) in $R$: (a) Cooper pair cotunneling; (b),(c) Cooper pair condensation; (d) Cooper pair tunneling. In the topological phase (e), (f) the direct tunneling via Majorana bound states contributes by (e) Cooper pair splitting and (f) quasiparticle tunneling.}
    \vspace{-0.5cm}
    \label{fig:process}
\end{figure}

We illustrate the processes contributing to gauge currents in this case in Fig.~\ref{fig:process}(a)-(d). For $\Delta_b = 0, w_b \neq 0$ we have the conventional Josephson junction and the current $j_L = - j_R = F_\text{t} \sin(2\phi_h)$ mediated by Cooper pair tunneling across the bridge. When $\Delta_b \neq 0, w_b= 0$, the currents $j_L = j_R = F_\text{c}\sin(2\phi_p)$ are mediated by Cooper pair cotunneling to or from the bridge. In the general case, $\Delta_b, w_b \neq 0$, the current $j_a$ contain cross terms $\sim \sin(\phi_p\pm\phi_h)$, contributed by Cooper pair condensation at the bridge. Remarkably, in this case the period of the Josephson current in $\phi_h$ and $\phi_p$ is doubled, similar to the junction with Majorana fermions and conserved fermion parity (see below)~\cite{Chiu_2019}.

In the topological (ordered) phase, the first-order contribution dominates since $H_b$ can now connect the unperturbed degenerate states. 
Then~\cite{SM},
\begin{align}
j_L^{(0)} &= P_0(F_\text{cs} \sin \phi_p + F_\text{qt} \sin \phi_h), \\
j_R^{(0)} &= P_0(F_\text{cs} \sin \phi_p - F_\text{qt} \sin \phi_h),
\end{align}
where $F_\text{cs} \propto \Delta_b$ and $F_\text{qt} \propto w_b$ are linear in the bridge tunneling amplitudes, and $P_0=\langle \Phi_0 | P |\Phi_0 \rangle$, where $P = \prod_j e^{i \pi n_j}$, is the parity of 
the state $\ket{\Phi_0}$ in the doubly degenerate manifold.
Note that $\epsilon_b / P_0 = -(F_\text{cs}\cos\phi_p + F_\text{qt}\cos\phi_h)$ switches sign in the cycle~\cite{SM}. The term $\sim \sin \phi_h$ is the fractional Josephson current mediated by Majorana quasiparticle tunneling across the bridge~\cite{Fu_2009}. The term $\sim \sin \phi_p$ is an unconventional Josephson current mediated by Cooper pair splitting at the bridge~\cite{Jiang_2011b}; see Figs.~\ref{fig:process}(e) and~\ref{fig:process}(f).

\begin{figure}[t]
\begin{center}
\includegraphics[width=3.2in]{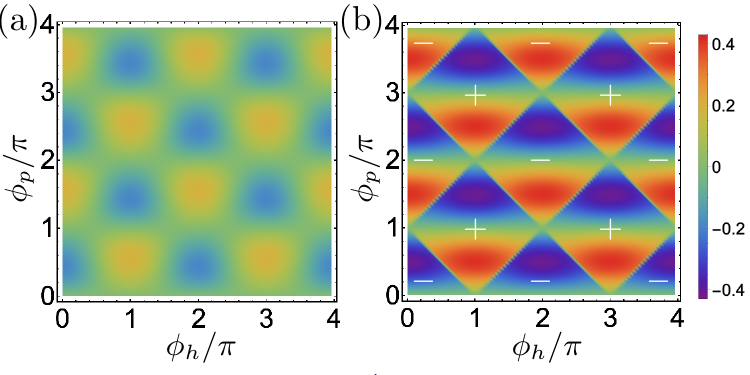}
\end{center}
\vspace{-0.7cm}
\caption{The gauge current $j_b = j_L + j_R$ through the bridge, scaled as $j_b/w_b^{2}$ in the trivial phase (a), and $j_b/w_b$ in the topological phase (b).
The parity of the ground state is fixed in the trivial phase (a), but switches sign in the topological phase (b). The parameters are 
$w_L = w_R = 1$ (units of energy), $\Delta_L = \Delta_R = 0.5, \Delta_{b} =  w_{b} = 0.2$; $\mu_L =\mu_R = 2.15$ in (a) and $\mu_L =\mu_R = 1.8$ in (b).
}
\vspace{-0.5cm}
\label{fig:fcurrent}
\end{figure}

The two phases of the spin system can thus be distinguished by the Floquet gauge current in two ways: (i) the linear (ordered) vs. quadratic (disordered) dependence of the the gauge current on the bridge tunneling amplitudes, and (ii) the dependence of the current on $\phi_h$ and $\phi_p$. The latter is often formulated as doubling of the periodicity of the current in the topological vs. trivial phase of fermions. This, in turn, relies on the conservation of the parity $P_0$ of the fermionic ground state. If, instead the system is prepared with the same sign of $\epsilon_b$, the parity of the ground state exhibits switches accompanied by discontinuities in the gauge current at topologically protected degeneracies in the topological phase, which are absent in the trivial phase. 

In terms of spins, the parity operator $P = \exp\left[{i\pi \sum_j (S^z_j + 1/2)}\right] = \prod_j (-2S^z_j)$. Therefore, the total fermion parity is the maximal multipoint spin-$z$ correlator. If the spin state is independently prepared for each value of $\phi_h$ and $\phi_p$, we should expect the state with the lowest energy is chosen. Therefore, the gauge current would show the same periodicity in both phases, while in the ordered phase it will show discontinuities accompanied with parity switches reflected in the sign reversals of the maximal multipoint spin-$z$ correlator.

\emph{Numerical results.}---%
To demonstrate these effects concretely, we have calculated the gauge current by exact diagonalization of the Floquet Hamiltonian in~(\ref{eq:HFa}) and~(\ref{eq:Hb}). In Fig.~\ref{fig:fcurrent}, we plot the average current $j_b^{(0)}=j_L^{(0)}+j_R^{(0)}$ through the bridge for a representative set of parameters realizing the trivial and topological phases of fermions as a function of gauge parameters $\phi_h$ and $\phi_p$. Here, we have chosen $\ket{\Phi_0}$ as the lowest energy state of the whole system. In the trivial phase, the current scales with $w_b^2$ and since both $w_b,\Delta_b\neq0$, compared to the conventional Josephson junction with $\Delta_b=0$, the periodicity in $\phi_p$ and $\phi_h$ is doubled. In the topological phase, the current scales with $w_b$ and its discontinuities coincide with parity switches, as expected. Note that except for discontinuous jumps due to parity switches in $P_0$, $j_b = 2P_0 F_\text{cs} \sin\phi_p$ has no other dependence on $\phi_h$.

\begin{figure}[t]
\begin{center}
\includegraphics[width=3.4in]{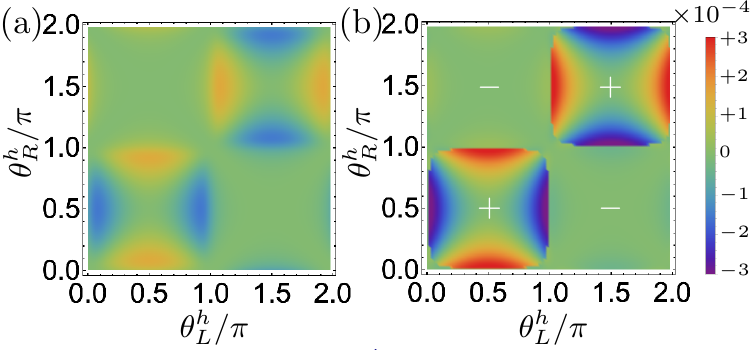}
\end{center}
\vspace{-0.7cm}
\caption{The magnetization current $d\braket{S^z_b}/dt$ in units of $\Omega$, at the bridge, $S^z_b=S^z_L+S^z_R$, for (a) disordered and (b) ordered phases of the driven XY model. The correlator $\braket{\prod_j(-2S^z_j)}$ is fixed in (a) and switches sign in (b). In units of $\Omega $, the parameters are $\overline J^x_L=\overline J^x_R=1.1\times10^{-4},\overline J^y_L=\overline J^y_R= 10^{-5}, \delta J^x_L=\delta J^x_R=3.2\times10^{-2}, \delta J^y_L=\delta J^y_R=2\times10^{-3},\theta^J= 0, \delta h^z_L= - \delta h^z_R= 0.03$.  In (a) $ \overline h_L^z=\overline h_R^z=3\times10^{-4} $; in (b) $ \overline h_L^z=\overline h_R^z=10^{-4}$.}
\vspace{-0.5cm}
\label{fig:scurrent}
\end{figure}

The magnetization current  of the Floquet gauge pump realized in the driven XY model is shown in Fig.~\ref{fig:scurrent} as a function of phase shifts $\theta_a^h-\theta^J$ across the bridge. Note that as these drive parameters are varied, both phases $\phi_h$ and $\phi_p$ as well as the amplitudes $w_b$, $\Delta_b$, $w_a$, and $\Delta_a$ change in a range determined by the drive amplitudes. We discuss this dependence in Supplemental Material~\cite{SM}. The difference between disordered [Fig.~\ref{fig:scurrent}(a)] and ordered [Fig.~\ref{fig:scurrent}(b)] phases is that there is a true discontinuous change in the current in Fig.~\ref{fig:scurrent}(b) while the change in Fig.~\ref{fig:scurrent}(a) is gradual. 
\highlight{The sign of the maximal multipoint spin-$z$ correlator provides a complementary signature of ground-state degeneracy.}


\emph{Experimental realization.}---%
The Floquet gauge pump can be realized using trapped atomic ions, which are a well-established system for simulating the time evolution of spin-lattice Hamiltonians \cite{monroe2019programmable}. Ions form defect-free lattices and can support quantum coherence times of longer than 10 minutes \cite{wang2017single}. Interactions between ions---which map to interactions between effective quantum spins---can be fully controlled and reprogrammed using laser light \cite{molmer1999multiparticle}. These features have made trapped ions the leading platform for establishing atomic frequency standards \cite{ludlow2015optical} and for performing quantum simulations of 1D spin chains.

All necessary components for implementing the Floquet gauge pump have been previously demonstrated in trapped-ion systems. Transverse-field Ising and XY models are routinely generated by driving stimulated Raman transitions between the effective spin states \cite{richerme2014non,jurcevic2014quasiparticle}. The resulting spin model depends upon the specific amplitude, frequency, and phase characteristics of the Raman laser beams, which can be controlled by passing the beams through an acousto-optic modulator (AOM) \cite{lee2016engineering}. Periodically driving the amplitude of the rf signal applied to the AOMs will imprint itself as a periodic drive on the spin-spin couplings $\{J^x(t),J^y(t)\}$ and transverse fields $h^z(t)$; indeed, this technique has already been used to realize Floquet engineering of a trapped-ion crystal \cite{zhang2017observation}. The one key challenge will be to implement asymmetric drives on the left and right halves of the chain. This may be solved by either (i) adding a second pair of Raman beams so that both halves can be independently addressed, or by (ii) adding a second pair of frequency components to the AOM rf drive to split a single Raman beam into two parts, each with its own amplitude, frequency, phase, and deflection angle.

\highlight{Topological degeneracies are detected by discontinuities in $d\braket{S^z(t)}/dt$. At high frequencies $\Omega \sim$~MHz in Fig.~\ref{fig:scurrent}, $d\braket{S^z(t)}/dt \sim 100$~Hz (see~\cite{SM}). Full contrast is obtained over $\sim 10$~ms. High resolution of 1 part in $1000$ in $\theta^h$ is also easily achieved.} A strength of trapped-ion systems is the ability to perform site-resolved spin-dependent fluorescence, which acts as a projective measurement along the $z$ direction and can discriminate between spin states with $> 99.9\%$ fidelity \cite{noek2013high}. Since all effective spins are detected simultaneously during the measurement, all possible spin correlators \highlight{(including the $N$-body correlator $\langle S_1^z S_2^z\ldots S_N^z \rangle$)} can be reconstructed when repeated trials are averaged. 

\emph{Concluding remarks.}---%
 Dynamical probes to interrogate and uncover the emergent discrete symmetries that give rise to ground-state degeneracies are key to discovering topological or broken-symmetry phases of matter. The Floquet gauge pump we have introduced in this Letter is a unique experimental tool conceiving this goal. 

Although our proposed proof-of-principle 
implementation in an ion-trap platform involved simple many-body systems, we expect 
that the Floquet gauge pump becomes a routine technique even in complex interacting 
systems. \highlight{The required ingredient is the existence of the gauge group. A sufficient condition in the spin models is the absence of couplings between the directions parallel and perpendicular to the field. The gauge group is then the rotations around the field, which relabels the axes in the perpendicular direction and leaves the spectrum unchanged.} In the XY model, for example, we can see immediately that the gauge pump works just as well in the presence of $S^z$-dependent interactions for spins~\cite{Herviou_2016,Mahyaeh_2020}, since these operators commute with the gauge current~\cite{SM}. In practical setups of ion-trap simulators, tunable, variable-range interactions between effective spin degrees of freedom are realized, \highlight{also in two-dimensional geometries~\cite{DOnofrio_2020}}, adding the possibility 
of magnetic frustration and potentially leading to exotic quantum spin liquid phases. Moreover, 
unveiling discrete symmetries can help in understanding the mechanisms leading to the formation 
of localized many-body boundary modes~\cite{Ortiz_2014,Cobanera_2014,Ortiz_2016}. This is of fundamental importance for 
practical applications, in particular, if the vacuum is topological and those modes represent 
quasiparticles with non-Abelian braiding statistics. 

\begin{acknowledgments}
This work is supported primarily by the U.S. Department of Energy, Office of Science, Basic Energy Sciences, under Award No. DE-SC0020343. B.S. and A.K. were supported in part by NSF CAREER Grant No. DMR-1350663 (early work on Floquet Hamiltonians).
\end{acknowledgments}

\bibliography{FGP-min-v2}


\vspace{-5mm}
\onecolumngrid
\newpage

\renewcommand{\thefigure}{S\arabic{figure}}
\renewcommand{\theequation}{S\arabic{equation}}
\setcounter{equation}{0}
\setcounter{figure}{0}

\title{
Supplemental Material for ``Floquet Gauge Pumps as Sensors for Spectral Degeneracies Protected by Symmetry or Topology''
}

\author{Abhishek Kumar}
\affiliation{Department of Physics, Indiana University, Bloomington, Indiana 47405, USA}

\author{Gerardo Ortiz}
\affiliation{Department of Physics, Indiana University, Bloomington, Indiana 47405, USA}
\affiliation{Quantum Science and Engineering Center, Indiana University, Bloomington, Indiana 47405, USA}

\author{Philip Richerme}
\affiliation{Department of Physics, Indiana University, Bloomington, Indiana 47405, USA}
\affiliation{Quantum Science and Engineering Center, Indiana University, Bloomington, Indiana 47405, USA}

\author{Babak Seradjeh}
\affiliation{Department of Physics, Indiana University, Bloomington, Indiana 47405, USA}
\affiliation{Quantum Science and Engineering Center, Indiana University, Bloomington, Indiana 47405, USA}
\affiliation{IU Center for Spacetime Symmetries, Indiana University, Bloomington, Indiana 47405, USA}

\begin{abstract}
Here we provide details of gauge current; Floquet gauge current; Floquet Hamiltonian of driven XY model; Time-averaged Floquet gauge current in the driven XY and equivalent fermion model; and additional numerical results.
\end{abstract}

\date{\today}

{
\let\clearpage\relax
\maketitle
}

\date{\today}

\onecolumngrid

\section{Gauge current}
The gauge currents were defined in the main text Eqs.~(1) and~(2) as $j_a := \braket{\partial H_\text{gp}/\partial\phi_a}$ on each side $a=L,R$. Here we confirm that these definitions agree with the usual fermionic currents
\begin{equation}
\frac{d}{dt}\braket{N_a} = -i\braket{[N_a, H_\text{gp}]},
\end{equation}
where the number operator $N_a = \sum_{j\in a} c_j^\dagger c\nodag_j$ and we have set $\hbar=1$. Note that for $j\in a$, $[N_a,c_j^\dagger] = c^\dagger_j$ and $[N_a,c_j] = - c_j$. In order to connect to the gauge structure of the Hamiltonian, we write these relations as
\begin{align}
[N_a,c_j^\dagger] 
	&= i\frac{\partial}{\partial \phi} \left[e^{-i\phi} c_j^\dagger\right]\bigg\vert_{\phi=0}, \\
[N_a, c_j] 
	&= i\frac{\partial}{\partial\phi}\left[ e^{i\phi} c_j \right]\bigg\vert_{\phi=0}.
\end{align}
So taking $H_\text{gp}$ to be normal ordered in $\{c_j^\dagger,c\nodag_j\}$, we have 
\begin{align}
[N_a,H_\text{gp}] 
	= i \frac{\partial \widetilde H_\text{gp}}{\partial \phi}\bigg\vert_{\phi=0},
\end{align}
where $\widetilde H_\text{gp}$ is obtained by replacing $c_{j}^\dagger \to e^{-i\phi}c_j^\dagger$ and $c_{j} \to e^{i\phi} c_j$ for $j\in a$. This is of course the U(1) fermion number gauge group. As a result, the angle $\phi$ adds to $\phi_a$, any gauge angle already present in the Hamiltonian. Therefore,
\begin{equation}
\frac{d}{dt}\braket{N_a} = \frac{\partial \braket{ \widetilde H_\text{gp}}}{\partial \phi}\bigg\vert_{\phi=0} =  \frac{\partial\braket{H_\text{gp}}}{\partial \phi_a} \equiv j_a.
\end{equation}

\section{Floquet gauge current} 
In this section we derive an expression for Floquet gauge current $j_a(t) = \braket{\partial H_\text{gp}(t)/\partial\phi_a}$. In a Floquet state $\ket{\Psi(t)} = e^{-i\epsilon t}\ket{\Phi (t)}$, where $\ket{\Phi(t)} = \ket{\Phi(t+2\pi/\Omega)}$ is periodic, normalized $\inner{\Phi(t)}{\Phi(t)}=1$, and satisfies the Floquet-Schr\"odinger equation
\begin{equation}
\left[ H_\text{gp}(t) - i \frac{\partial}{\partial t} \right]\ket{\Phi(t)} = \epsilon \ket{\Phi(t)},
\end{equation}
with quasienergy $\epsilon$, the current
$
j_a (t) = \bra{\Phi(t)} {\partial H_\text{gp}(t)}/{\partial \phi_a} \ket{\Phi(t)},
$
is also periodic and can be expanded in Fourier modes. Taking a derivative $\partial/\partial\phi_a$ of the Floquet-Schr\"odinger equation, we find a Floquet version of the Hellman-Feynman theorem,
\begin{align}
j_a(t) = \bra{\Phi(t)}\frac{\partial H_\text{gp}(t)}{\partial \phi_\alpha} \ket{\Phi(t)} = i \frac{\partial}{\partial t} \bra{\Phi(t)}\frac{\partial }{\partial \phi_\alpha} \ket{\Phi(t)}+\frac{\partial \epsilon}{\partial \phi_\alpha}.
\end{align}
Using the Fourier expansion $ \ket{\Phi(t)}= \sum_m \ket{\Phi^{(m)}} e^{-im\Omega t}$, we find the Fourier components of the gauge current as
\begin{align}
j_a^{(0)} \label{eq:ja0}
	&= \frac{\partial \epsilon}{\partial \phi_a}, \\
j_a^{(n\neq 0)} 
	&= n\Omega \sum_{m\in\mathbb{Z}}\bra{\Phi^{(m)}}\frac{\partial}{\partial \phi_a}\ket{\Phi^{(m+n)}}.
\end{align}
Eq.~(\ref{eq:ja0}) is the same as Eq.~(4) in the main text.
 
\begin{figure}[t]
\begin{center}
\includegraphics[width=5in]{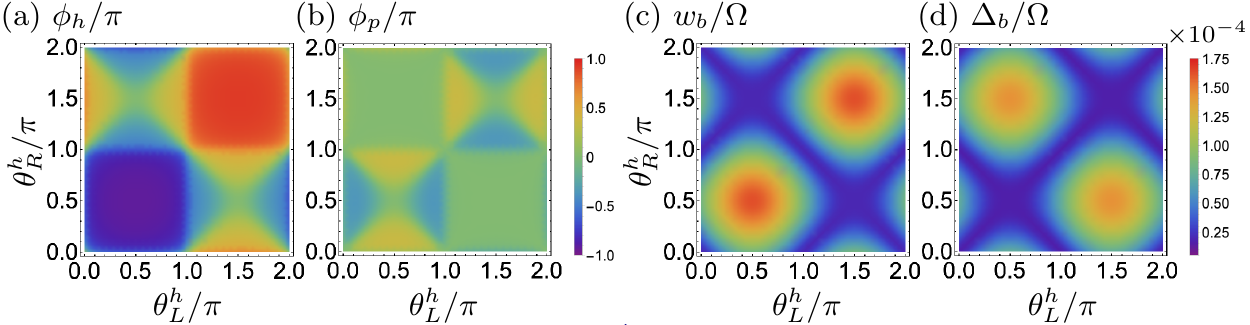}
\end{center}
\caption{The bridge parameters, (a) hopping phase $\phi_h$, (b) pairing phase $\phi_p$, (c) hopping amplitude $w_b$, and (d) paring amplitude $\Delta_b$ as a function of the drive parameters $\theta^h_L$ and $\theta^h_R$. The values of other parameters are $\overline J^x=1.1\times10^{-4}, \overline J^y= 10^{-5}, \delta J^x_L = \delta J^x_R = 3.2\times10^{-2}, \delta J^y_L=\delta J^y_R=2\times10^{-3}, h^z=\times10^{-4}, \delta h^z_L= -\delta h^z_R= 3\times10^{-2}$ (in units of $\Omega$) and $\theta^J= 0$, as in Fig. 4 of the main text. 
}
\label{fig:S1}
\end{figure}
 
\section{Floquet Hamiltonian of driven XY model}
Starting with the driven XY hamiltonian
\begin{align}
    H_\text{XY}(t)= \sum_{j} \left[ J_j^{x}(t)S^{x}_j S^{x}_{j+1} +J_j^{y}(t) S^{y}_j S^{y}_{j+1} + h_j^{z}(t) S^{z}_j \right], 
\end{align}
with $J_{j}^{x,y}(t) = \overline J_j^{x,y} + \delta J^{x,y}_{j} \cos(\Omega t +\theta^J) $ and $h^{z}_{j}(t) = \overline h_j^z +  \delta h^{z}_{j} \cos(\Omega t +\theta^h_j)$, we have the Fourier components
\begin{align}
H^{(0)} &=  \sum_{j}\big( \overline{J}_j^{x} S^{x}_j S^{x}_{j+1} +\overline{J}_j^{y} S^{y}_j S^{y}_{j+1} + \overline{h}^{z}_{j} S^{z}_j), \\
H^{(1)} &= {H^{(-1)}}^{\dagger} = \frac12 \sum_{j}\left[ \left( {\delta J_j^{x} } S^{x}_j S^{x}_{j+1} + {\delta J_j^{y}} S^{y}_j S^{y}_{j+1} \right)e^{-i\theta^J} + {\delta h^{z}_{j} } e^{-i\theta^h_j}S^{z}_j \right].
\end{align}
Now, using the commutation relations $[S_j^\alpha,S_l^\beta] = i \delta_{jl} \epsilon^{\alpha\beta\kappa}S_j^{\kappa}$, we find after some algebra
\begin{align}
H_F 
	&= H^{(0)} +  \sum_{n > 0} \frac{[H^{(-n)},H^{(n)}]}{n\Omega} = H^{(0)} +  \frac{[H^{(-1)},H^{(1)}]}{\Omega} \\ 
	&= H^{(0)} + \frac{1}{2\Omega}\sum_{j}\left[  \delta J_j^{+} \left( -\tilde{\delta h}_{j+1}^z + \tilde{\delta h}_{j}^z  \right) \left(S^{x}_j S^{y}_{j+1}-S^{y}_j S^{x}_{j+1}\right) - \delta J_j^{-}\left( \tilde{\delta h}_{j+1}^z + \tilde{\delta h}_{j}^z \right)  \left(S^{x}_j S^{y}_{j+1}+S^{y}_j S^{x}_{j+1}\right) \right]  
\end{align}
where we have defined $\tilde{\delta h}_{j}^z =\delta h_{j}^z\sin(\theta_{j}^h-\theta^J)$ and $\delta J_j^{\pm}= \frac12 (\delta J_j^{x} \pm  \delta J_j^{y})$. The first term in the bracket is the local Dzyaloshinsky-Moriya Interaction (DMI) generated dynamically by the drive. Now, taking the drive parameters to be uniform on a given side $a$, $\overline J^{x,y}_{j\in a} = \overline J^{x,y}_a$, $\delta J^{x,y}_{j\in a} = \delta J^{x,y}_a$, etc. we find Eqs.~(5) and~(6) of the main text. 

In Fig.~\ref{fig:S1}, we show the dependence of the pairing and hopping terms at the bridge on the drive parameters.

\section{Time-averaged Floquet gauge current}
Here we will calculate the time-averaged Floquet gauge current. For this we have use the expression derived in Eq.~\eqref{eq:ja0}. Switching to bridge variables $\phi_h = \phi_L-\phi_R$ and $\phi_p = \phi_L + \phi_R$ (up to constant phases in the bridge),
\begin{align}
j^{(0)}_L =  \frac{\partial \epsilon}{\partial \phi_p} + \frac{\partial \epsilon}{\partial \phi_h}, \quad
j^{(0)}_R =  \frac{\partial \epsilon}{\partial \phi_p} - \frac{\partial \epsilon}{\partial \phi_h}.
\end{align}
Here, $\epsilon$ is the quasienergy of the full Floquet Hamiltonian (after the gauge transformation) $\widetilde H_L^\text{F} +  H_{b}(\phi_h,\phi_p)+ \widetilde H_R^\text{F}$
\begin{align}
\widetilde H_{a}^\text{F} &= \sum_{j\in a} \left[ w_{a} c_j^{\dagger}c\nodag_{j+1} + \Delta_{a} c_j^{\dagger}c_{j+1}^{\dagger} + \mu_a n_j \right] + \text{H.c.}\\
H_{b} &=  w_{b}e^{i\phi_h} c^{\dagger}_{\mycal{l}} c\nodag_{\mycal{r}} +  \Delta_{b}e^{i\phi_p} c^{\dagger}_{\mycal{l}} c^{\dagger}_{\mycal{r}} + \text{H.c.}
\end{align}
in the fermion basis, with real parameters $w_a,\Delta_a,w_b$, and $\Delta_b$.
We will use perturbation theory with $H_0 = \widetilde H_L^\text{F} + \widetilde H_R^\text{F}$ as the unperturbed and $H_{b}(\phi_h,\phi_p)$ as perturbation Hamiltonians.  An unperturbed Floquet state $\ket{\Phi_0}$ of $H_0$ and its quasienergy $\epsilon_0$ are independent of $\phi_h,\phi_p$. The quasienergy up to second-order is $\epsilon = \epsilon_0 + \epsilon_b$ with
\begin{equation}
\epsilon_b(\phi_h,\phi_p) =  \bra{ \Phi_0} H_{b} \ket{\Phi_0} + \sum_{\alpha\neq0}\frac{|\bra{\Phi_0}H_{b} \ket{\Phi_\alpha}|^{2}}{\epsilon_0 - \epsilon_{\alpha}},
\end{equation}
where $\ket{\Phi_{\alpha}}$ are eigenstates of $H_0$ with quasienergy $\epsilon_\alpha$. 
All the dependence on $\phi_h$ and $\phi_p$ is contained in $H_b$.

Performing a mode expansion,
\begin{align}
c^{\dagger}_{\mycal{l}} &= \sum_{\epsilon_L > 0} \left( u^{*}_{\epsilon_L,\mycal{l}} \gamma^{\dagger}_{\epsilon_L} + v_{\epsilon_L,\mycal{l}} \gamma_{\epsilon_L} \right), \\
c^{\dagger}_{\mycal{r}} &= \sum_{\epsilon_R > 0} \left( u^{*}_{\epsilon_R,\mycal{l}} \gamma^{\dagger}_{\epsilon_R} + v_{\epsilon_R,\mycal{r}} \gamma_{\epsilon_R}\right),
\end{align}
in terms of Bogoliubov operators $\gamma_{\epsilon_a}$ that form the diagonal basis of $\widetilde H_a^\text{F}$ with particle (hole) wavefunctions $u_{\epsilon_a,j\in a}$ ($v_{\epsilon_a,j\in a}$) and single-body excitation quasienergy $\epsilon_a$. We choose the state $\ket{\Phi_0} = \ket{\Phi_{0L}}\otimes\ket{\Phi_{0R}}$ such that
\begin{equation}
\gamma_{\epsilon_a}\ket{\Phi_{0a}} = 0,
\end{equation}
for all values $\epsilon_{a}>0$. When there are zero-energy solutions, the state is degenerate; we will deal with this case separately below. This choice is consistent with the Floquet state in the high-frequency limit obtained continuously from the ground state of the time-average Hamiltonian.

We can now expand $H_b$ and calculate the overlaps to find $\epsilon$.

\subsection*{Trivial (nondegenerate) case}
In this case, the first order perturbative energy is zero since all terms like $\bra{\Phi_0} \gamma_{\epsilon_L}\gamma_{\epsilon_R} \ket{\Phi_0}$, $\bra{\Phi_0} \gamma_{\epsilon_L}\gamma_{\epsilon_R}^\dagger \ket{\Phi_0}$, etc. vanish. In the second order, all overlaps involving $\gamma_{\epsilon_a}^\dagger$ vanish since $\bra{\Phi_0}\gamma_{\epsilon_a}^\dagger = 0$.
After some algebra, we find that the only terms contributing are
\begin{align}
\bra{\Phi_0} e^{i\phi_h}c^{\dagger}_{\mycal{l}} c\nodag_{\mycal{r}} + \text{H.c.} \ket{\Phi_\alpha} 
	&= \sum_{\epsilon_L,\epsilon_R > 0} \left( e^{i\phi_h} z_{LR}^{+-} - e^{-i\phi_h} z_{LR}^{-+} \right) \bra{\Phi_0}\gamma_{\epsilon_L} \gamma_{\epsilon_R}\ket{\Phi_\alpha}, \\
\bra{\Phi_0} e^{i\phi_p}c^{\dagger}_{\mycal{l}} c^\dagger_{\mycal{r}} + \text{H.c.} \ket{\Phi_\alpha} 
	&= \sum_{\epsilon_L,\epsilon_R > 0} \left( e^{i\phi_p} z_{LR}^{++} - e^{-i\phi_p} z_{LR}^{--} \right) \bra{\Phi_0}\gamma_{\epsilon_L} \gamma_{\epsilon_R}\ket{\Phi_\alpha}.
\end{align}
where the coherence factors
\begin{align}
z_{LR}^{+-} = v_{\epsilon_L,{\mycal{l}}} u_{\epsilon_R,{\mycal{r}}},  \quad
z_{LR}^{-+} = u_{\epsilon_L,{\mycal{l}}} v_{\epsilon_R,{\mycal{r}}}, \quad
z_{LR}^{++} = v_{\epsilon_L,{\mycal{l}}} v_{\epsilon_R,{\mycal{r}}}, \quad
z_{LR}^{--} = u_{\epsilon_L,{\mycal{l}}} u_{\epsilon_R,{\mycal{r}}}.
\end{align}
The remaining overlap $\bra{\Phi_0}\gamma_{\epsilon_L} \gamma_{\epsilon_R}\ket{\Phi_\alpha} = 1$ when $\epsilon_\alpha - \epsilon_0 = \epsilon_L + \epsilon_R$ and vanishes otherwise.

Since $H_0$ is entirely real (complex phases are gauged away), we can choose the wavefunctions $u_{\epsilon_a}, v_{\epsilon_a}$ to be real. Then, after some algebra, we find
\begin{equation}
 \epsilon = -F_0 - \frac12 F_\text{t}\cos(2\phi_h) - \frac12 F_\text{c}\cos(2\phi_p) - \frac12 F_L\cos(\phi_p+\phi_h) - \frac12 F_R\cos(\phi_p-\phi_h),
\end{equation}
with
\begin{align}
F_0 
	&= \sum_{\epsilon_L,\epsilon_R > 0} \frac{w_b^{2} \left( |z_{LR}^{+-}|^{2}  + |z_{LR}^{-+}|^{2}\right)+ \Delta_b^2\left( |z_{LR}^{++}|^{2}  + |z_{LR}^{--}|^{2}\right)}{\epsilon_L+\epsilon_R}, \\
F_\text{t} 
	&= -4w_b^{2} \sum_{\epsilon_L,\epsilon_R > 0} \frac{z_{LR}^{+-}z_{LR}^{-+}}{\epsilon_L+\epsilon_R}, \\
F_\text{c}
	&= -4\Delta_b^{2} \sum_{\epsilon_L,\epsilon_R > 0} \frac{z_{LR}^{++}z_{LR}^{--}}{\epsilon_L+\epsilon_R}, \\
F_L 
	&= 4w_b\Delta_b \sum_{\epsilon_L,\epsilon_R > 0} \frac{z_{LR}^{+-}z_{LR}^{++} + z_{LR}^{-+}z_{LR}^{--}}{\epsilon_L+\epsilon_R},\\
F_R 
	&= 4w_b\Delta_b \sum_{\epsilon_L,\epsilon_R > 0} \frac{z_{LR}^{+-}z_{LR}^{--} + z_{LR}^{-+}z_{LR}^{++}}{\epsilon_L+\epsilon_R}.
\end{align}
Thus, Eqs.~(11) and~(12) of the main text follow.

\subsection*{Topological (degenerate) case}
In this case the states of the system are (at least) doubly degenerate. This is reflected in the presence of zero-energy Majorana modes at the ends of each side of the gauge pump Hamiltonian. The bridge Hamiltonian now has terms that can connect the degenerate states, so the leading contributions to the  quasienergy, $\epsilon_b$, in degenerate perturbation theory come from the linear first-order term. Labeling the Majorana modes at the junction as $\gamma_{L0}$ and $\gamma_{R0}$, we can take the two degenerate states $\ket{\Phi_0^\pm}$ to have opposite parities $i\gamma_{L0}\gamma_{R0}\ket{\Phi_0^\pm} = \pm \ket{\Phi_0^\pm}$. Since the parity of the other modes remains unchanged, these states have opposite total parity $\bra{\Phi_0^\pm} P \ket{\Phi_0^\pm}$ with $P = \prod_j e^{i\pi n_j}$.

For Majorana zero-modes $u_{0a}^*=v_{0a}$ with the two localized on the left and right end of the chain having orthogonal particle-hole spinors. One can choose, say, $u_{0L}\in\mathbb{R}$ and $u_{0R} = i \overline u_{0R}$ with $\overline u_{0R}\in\mathbb{R}$. Then,
\begin{align}
c^{\dagger}_{\mycal{l}} &= +2 u_{0L,\mycal{l}} \gamma_{0L} + \sum_{\epsilon_L > 0} \left( u^{*}_{\epsilon_L,\mycal{l}} \gamma^{\dagger}_{\epsilon_L} + v_{\epsilon_L,\mycal{l}} \gamma_{\epsilon_L} \right), \\
c^{\dagger}_{\mycal{r}} &= -2 i \overline u_{0R,\mycal{r}} \gamma_{0R} + \sum_{\epsilon_R > 0} \left( u^{*}_{\epsilon_R,\mycal{l}} \gamma^{\dagger}_{\epsilon_R} + v_{\epsilon_R,\mycal{r}} \gamma_{\epsilon_R}\right),
\end{align}
We can now see that the non-zero contributions to $\epsilon_b$ are of the form
\begin{align}
\bra{\Phi_0} e^{i\phi_h}c^{\dagger}_{\mycal{l}} c\nodag_{\mycal{r}} + \text{H.c.} \ket{\Phi_0} 
	&= +8u_{0L,\mycal{l}} \overline u_{0R,\mycal{r}}  \bra{\Phi_0} i \gamma_{0L}\gamma_{0R} \ket{\Phi_0} \cos\phi_h, \\
\bra{\Phi_0} e^{i\phi_p}c^{\dagger}_{\mycal{l}} c^\dagger_{\mycal{r}} + \text{H.c.} \ket{\Phi_0} 
	&= -8u_{0L,\mycal{l}} \overline u_{0R,\mycal{r}} \bra{\Phi_0} i \gamma_{0L}\gamma_{0R} \ket{\Phi_0} \cos\phi_p,
\end{align}
for a $\ket{\Phi_0}$ that is any linear combination of $\ket{\Phi_0^\pm}$.

Finally, the corrections to quasienergy of the state $\ket{\Phi_0}$ can be written as
\begin{align}
\epsilon_b = - P_0 \left( F_{\text{qt}} \cos\phi_h + F_{\text{cs}} \cos\phi_p \right),
\end{align}
with parity $P_0 = \bra{\Phi_0} i\gamma_{0L}\gamma_{0R} \ket{\Phi_0} = \bra{\Phi_0} P \ket{\Phi_0}$, and
\begin{align}\label{eq:Fqc}
F_{\text{qt}} = -8u_{0L,\mycal{l}} \overline u_{0R,\mycal{r}}  w_b, \quad
F_{\text{cs}} = +8u_{0L,\mycal{l}} \overline u_{0R,\mycal{r}} \Delta_b.
\end{align}
The gauge currents given in Eqs.~(13) and~(14) then follow straightforwardly.

\section{Additional numerical results}
\begin{figure}[bt]
\begin{center}
\includegraphics[width=5in]{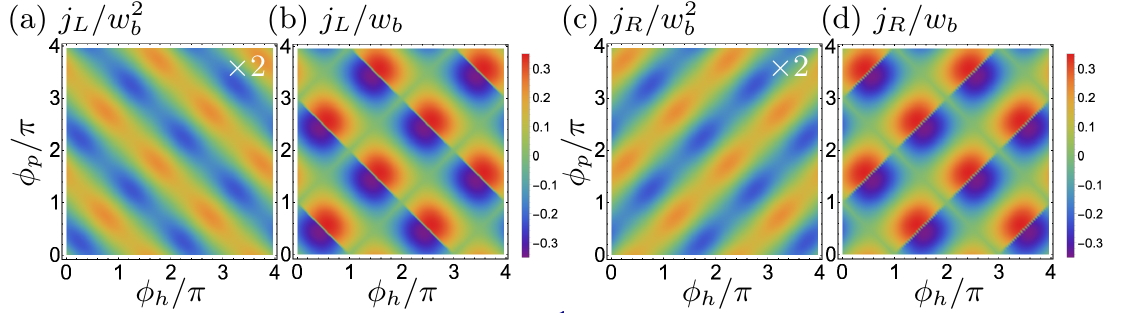}
\end{center}
\vspace{-3mm}
\caption{Gauge currents trivial (a,c) and topological (b,d) phases on each side. The parameters are  $w_L = w_R = 1$ (units of energy), $\Delta_L = \Delta_R = 0.5, \Delta_{b} =  w_{b} = 0.2$; $\mu_L =\mu_R = 2.15$ in (a),(c) and $\mu_L =\mu_R = 1.8$ in (b),(d).}
\label{fig:S2}
\end{figure}

Here we show some additional numerical results for the gauge current. 
In Fig.~\ref{fig:S2}, we plot the gauge current $j_a$ on each side in both the trivial and topological phases for the case with $w_b\neq0=\Delta\neq0$ as in Fig.~3 of the main text. The currents show smooth behavior in the trivial phase and jumps in the topological phase. However, the pattern of jumps is now associated with parity switches in different parts of the system, not the total parity.

\begin{figure}[bt]
\begin{center}
\includegraphics[width=5in]{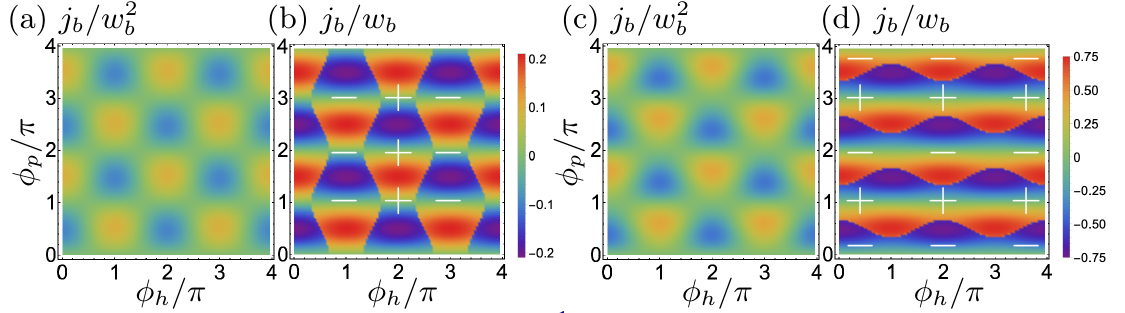}
\end{center}
\vspace{-3mm}
\caption{The gauge current $j_b = j_L + j_R$ through the bridge in the trivial phase (a,c) and in the topological phase (b,d).
The total parity is fixed in the trivial phase, but switches sign in the topological phase. The parameters are  $w_L = w_R = 1$ (units of energy), $\Delta_L = \Delta_R = 0.5, \Delta_{b} =  w_{b} = 0.2$; $\mu_L =\mu_R = 2.15$ in (a,c) and $\mu_L =\mu_R = 1.8$ in (b,d); $\Delta_{b} =0.1,  w_{b} = 0.2$ in (a,c) and $\Delta_{b} =0.2,  w_{b} = 0.1$ in (b,d).}
\label{fig:S3}
\end{figure}

In Fig.~\ref{fig:S3}, we plot the gauge current $j_b$ at the bridge, like in Fig.~3 of the main text, but with $w_b\neq\Delta_b$. Now, the pattern of parity switches in the topological phase is changed depending on whether $w_b>\Delta_b$ (panels b) or $w_b<\Delta_b$ (panel d). This is consistent with our analytical calculation in Eq.~(\ref{eq:Fqc}) showing $|F_\text{qt}/F_\text{cs}| = |w_b/\Delta_b|$.

\begin{figure}[bt]
\begin{center}
\includegraphics[width=5.5in]{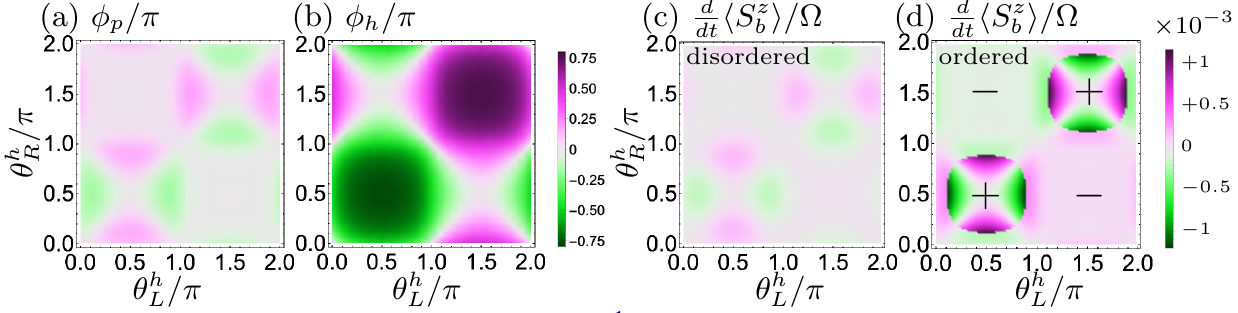}
\end{center}
\vspace{-3mm}
\caption{The bridge gauge parameters (a,b) and the bridge magnetization current in the disordered (c) and ordered (d) phases as a function of drive parameters $\theta_L^h$ and $\theta_R^h$. The maximal multipoint spin-$z$ correlator $\braket{\prod_j(-2S^z_j)}$ is fixed in (c) and switches sign $\pm$ in (d) as shown. In units of $\Omega $, the parameters are $\overline J^x_L=\overline J^x_R=1.1\times10^{-3},\overline J^y_L=\overline J^y_R= 10^{-4}, \delta J^x_L=3.5\times10^{-2}, \delta J^x_R=3\times10^{-2}, \delta J^y_L=\delta J^y_R=2\times10^{-3}, \delta h^z_L= - \delta h^z_R= 0.09$, and $\theta^J= 0$.  In (c) $ \overline h_L^z=\overline h_R^z=3\times10^{-3} $; in (d) $ \overline h_L^z=\overline h_R^z=10^{-3}$.}
\label{fig:S4}
\end{figure}

Finally, in Fig.~\ref{fig:S4}, we show a choice of parameters for the ion trap realization of the Floquet gauge pump in the high-frequency regime, for which the gauge current is $\sim 10^{-3}\Omega$, an order of magnitude larger than in Fig. 4. As discussed in the main text, this makes it easier to detect the discontinuities in the gauge current.


\end{document}